\documentstyle[12pt]{article}
\def\ie{i.\ e.\ }
\def\etal{et al.\ }
\def\lsim{\stackrel{<}{\sim}}

\begin{document}

\baselineskip 18pt

\newcommand{\sheptitle}
{Charginos and Neutralinos at LEPII}

\newcommand{\shepauthor}
{Marco A. D\'\i az and Steve F. King }

\newcommand{\shepaddress}
{Physics Department, University of Southampton\\
Southampton, SO17 1BJ, U.K.}

\newcommand{\shepabstract}
{We show that LEPII will either discover charginos and neutralinos,
or enable a very stringent upper limit to be placed
on $\tan \beta$ as a function of the gluino mass.
The only assumption we make is the existence of some
unified model which breaks down to the
minimal supersymmetric standard model below the unification scale.
In such a framework we discuss how the discovery of a chargino at LEPII
and the measurement of its mass and production cross-section,
together with the measurement of the mass of the lightest
neutralino, would enable the gluino mass, $\tan \beta$
and $\mu$ to be predicted, up to a possible ambiguity in the
sign of $\mu$ which we discuss.}

\begin{titlepage}
\begin{flushright}
SHEP 95-01\\
January 1995 \\
\end{flushright}
\vspace{.4in}
\begin{center}
{\large{\bf \sheptitle}}
\bigskip \\ \shepauthor \\ \mbox{} \\ {\it \shepaddress} \\ \vspace{.5in}
{\bf Abstract} \bigskip \end{center} \setcounter{page}{0}
\shepabstract
\end{titlepage}

\newpage
The LEP $e^+e^-$ collider
provides a clean environment for searching for the charginos and
neutralinos predicted by the
minimal supersymmetric standard model (MSSM) \cite{MSSMrep}.
In the context of a unified model it may be assumed
that the three low energy gaugino mass parameters $M_i$,
with $i=1,2,3$ corresponding to the groups $U(1)$, $SU(2)$, and
$SU(3)$ respectively,
are unified into a universal gaugino mass $M_{1/2}$ at the same
scale $M_X$ at which the gauge couplings $\alpha_i=g_i^2/{4\pi}$
are unified into a single coupling $\alpha_X$ \cite{alphauni}.
Then at the electroweak scale (which we take to be $m_Z$),
\begin{equation}
\frac{M_i}{M_{1/2}}=\frac{\alpha_i(m_Z^2)}{\alpha_X}
\end{equation}
as a simple consequence of the one-loop renormalisation group
equations. For example the gluino mass is given by
\begin{equation}
m_{\tilde{g}}=M_3={M_{1/2}}\frac{\alpha_3(m_Z^2)}{\alpha_X}.
\end{equation}
Having made this assumption the chargino
$\tilde{\chi}^{\pm}_i$ ($i=1,2$)
and neutralino $\tilde{\chi}^0_i$ ($i=1 \ldots 4$) masses
and mixing angles then only depend on three unknown parameters:
the gluino mass $m_{\tilde{g}}$, $\mu$ and $\tan \beta$
\cite{GunionHaber}.

Several authors have considered the production of neutralinos
and charginos at hadron colliders \cite{prodHad},
$e^+e^-$ Colliders at the $Z$ pole \cite{prodLEP}
and beyond \cite{prodLEPII}, as well as its
decay modes \cite{decays}.
The results of such analyses are usually presented
as allowed and excluded regions in the $m_{\tilde{g}}-\mu$
plane for specified values of $\tan \beta$ \cite{Baer}.
Given the already widespread interest in this subject,
we should be clear to point out what we have done that has
not already been considered.
The purpose of the present paper is twofold:

(1) We shall show how existing LEP data may be used to
extract a precise bound on $\tan \beta$ as a function
of $m_{\tilde{g}}$.
In order to do this we shall
present our results in the $\tan \beta-\mu$
plane for specified values of $m_{\tilde{g}}$.
This enables us to deduce an
upper bound on $\tan \beta$ as a function of $m_{\tilde{g}}$.
Since the bound is monotonic, the maximum value of $\tan \beta$
is equivalent to specifying the
minimum value of $m_{\tilde{g}}$ for a given value of $\tan \beta$,
and this information is available from the $\tan \beta =1,2,5,30$
contours in the $m_{\tilde{g}}-\mu$ plane \cite{Baer}.
However our method
makes it possible to
extract a precise bound on $\tan \beta$ as a function
of $m_{\tilde{g}}$, at LEP.

(2) We shall consider the prospects for chargino
discovery at LEPII. Either the chargino will not
be discovered, in which case we show that this leads to
a very stringent bound on $\tan \beta$ as a function
of $m_{\tilde{g}}$. Or the lighter chargino and (by virtue of its decay)
the lightest neutralino will be discovered
in which case we show how this would
enable the gluino mass, $\tan \beta$
and $\mu$ to be predicted from the LEPII measurements
of the mass of the chargino, the mass of the lightest neutralino,
and the measurement of the chargino production cross-section.
We try to present these numerical
results in a form which will be useful
to experimentalists.

We begin by considering the constraints on
the MSSM coming from the negative searches at LEP for new
particles with the signatures of
charginos and neutralinos. For gluino masses not too heavy,
the allowed region of the $\tan \beta$ plane corresponds to
negative values of $\mu$
\footnote{ We use the same convention as \cite{Baer}
in which the superpotential $W=-\mu H_1H_2$
where $H_1H_2=H_1^0H_2^0-H_1^-H_2^+$.}
and in practice is determined by the intersection
of the regions allowed by the following two constraints:

\begin{itemize}

\item{The 95\% CL upper bound on the contribution of new particles
to the Z width is
$\Delta \Gamma_Z < 30$ MeV.}

\item{The branching fraction for the decays $Z\rightarrow
\tilde{\chi}^0_i \tilde{\chi}^0_j$ (where i and j are not both 1)
satisfies
\begin{equation}
B(Z\rightarrow \tilde{\chi}^0_i \tilde{\chi}^0_j)<10^{-5}.
\end{equation}  }
\end{itemize}

The first of these experimental
constraints is similar to that used previously
\cite{Baer}, and essentially is derived from the measurement of the
Z width. This constraint has a similar effect to
the constraint that $m_{\tilde\chi^{\pm}_1}>45$ GeV,
( at least for large negative regions of $\mu$) but
is always more restrictive for our range of gluino masses.
The second constraint above is
a factor of 5 more stringent than previously assumed \cite{Baer}.
and is based on
the negative searches for typical neutralino signatures
such as the preliminary OPAL result \cite{Glasgow}
\begin{equation}
B(Z\rightarrow \gamma X_{inv})<4.3 \times 10^{-6}
\end{equation}
for $M_{X_{inv}}<64$ GeV (95\% CL).
For our range of gluino mass, this constraint is always more
restrictive than the constraint that the invisible width
of the Z satisfies $\Delta \Gamma_{Z_{inv}} < 7$ MeV.
Constraints such as those mentioned above may be combined
by the LEP experiments to produce an excluded region in the
$m_{\tilde{\chi}^0_i}-\tan \beta$ plane. For example a preliminary
result from L3 gives $m_{\tilde{\chi}^0_1}>25$ GeV
and $m_{\tilde{\chi}^0_2}>42$ GeV for $\tan \beta>2$ \cite{Glasgow}.
We have checked that such excluded regions in the
$m_{\tilde{\chi}^0_i}-\tan \beta$ plane do not provide any additional
constraints in our analysis, for a sample value of the gluino mass.

In Fig.\ 1 we show in detail how the above LEP constraints
restrict the parameter space of the MSSM. For a gluino
mass of 140 GeV, Fig.\ 1 shows that
much of the $\tan \beta-\mu$ plane is excluded
by these constraints. For example all of the parameter space
with $\mu >0$ (mostly not shown) is excluded in this case.
In the negative $\mu$ region shown in Fig.\ 1 it is clearly seen
that there is a maximum value of $\tan \beta<5$
given by the intersection of the lines corresponding
to the constraints
$\Delta \Gamma_Z < 30$ MeV and
$B(Z\rightarrow \tilde{\chi}^0_i \tilde{\chi}^0_j)<10^{-5}$.
The other constraints mentioned above are shown for completeness
in Fig.\ 1 but do not provide additional restrictions
on the allowed region of the $\tan \beta-\mu$ plane.
However if LEPII fails to discover the lighter chargino
and sets a limit of $m_{\tilde{\chi}^{\pm}_1}>80$ GeV,
then the effect dramatically reduces the allowed region
in this plane to the very small region indicated near
$\tan \beta \approx 1$ and $\mu \approx -80$ GeV.

{}From analysing plots of the kind shown in Fig.\ 1 for
various gluino masses we are able to obtain
a rather precise
upper bound on $\tan \beta$ as a function of gluino mass
from existing LEP data, and a corresponding
but much more restrictive bound on $\tan \beta$ from the
assumption that LEPII will not discover the chargino,
and will set a limit of $m_{\tilde{\chi}^{\pm}_1}>80$ GeV.
Both these bounds are shown in Fig.\ 2. The model independent
bound on the gluino mass from CDF is $m_{\tilde{g}}> 100$ GeV
(90 \% c.l.) \cite{PDG}. \footnote{This bound allows the squarks to
be heavier than the gluino, and allows cascade decays \cite{Cascade}.}
These plots show that, should an intermediate mass gluino be discovered
at the Tevatron, then a precise and useful bound may be
placed upon $\tan \beta$ by LEPI or LEPII data, on the assumption
that no charginos or neutralinos are discovered at LEPII.

A light gluino is not excluded by experiments \cite{LightGluExp}.
The allowed light gluino window is $2.6<m_{\tilde g}\lsim 6$ GeV
and $m_{\tilde g}<0.6$ GeV, although the exact boundary of this
window is controversial \cite{window}. With the assumptions
presented in this paper, \ie, unification of gaugino masses,
there is an allowed region in the $\tan\beta-\mu$ plane
\cite{LightGluTheo,DiazGlu}. Nevertheless, in supergravity
models with a radiatively broken electroweak symmetry group
and universality of scalar and gaugino masses at the unification
scale the light gluino window has been closed \cite{DiazGlu}.

Now let us consider the possibility that LEPII will discover
the lighter chargino with a mass in the range
$m_{\tilde{\chi}^{\pm}_1}=50-90$ GeV. A typical signature
of chargino pair production would be a charged lepton pair
$l^+l^-$ plus missing energy,
arising from the decay $\tilde{\chi}^\pm_1
\rightarrow \tilde{\chi}^0_1W^{\pm \ast}$,
$W^{\pm \ast}\rightarrow l^{\pm} \nu$. Such a clean signature should
enable the chargino to be discovered right up to the kinematic limit
of the collider, and by scanning in energy the chargino mass should
be easily extracted. Clearly if LEPII discovers the lighter chargino then
it will simultaneously also discover the lightest neutralino into
which the chargino decays thereby discovering two new particles
for the price of one! By various kinematic means, such as
measuring the maximum charged lepton momenta, it should be
possible to measure the lightest neutralino mass from such events.
Given accurate measurements of $m_{\tilde{\chi}^{\pm}_1}$,
$m_{\tilde{\chi}^{0}_1}$ and the total cross-section
$\sigma (e^+e^-\rightarrow \tilde{\chi}^+_1 \tilde{\chi}^-_1)$ it will
be possible for LEPII to pin down all the parameters
of the chargino/neutralino sector, namely
$m_{\tilde{g}}$, $\tan \beta$ and $\mu$,
from which all the information about the entire chargino/neutralino
spectrum may be determined, assuming unification.

In Figs.\ 3-7 we show gluino mass contours
in the $\sigma (e^+e^-\rightarrow \tilde{\chi}^+_1
\tilde{\chi}^-_1)-m_{\tilde{\chi}^{0}_1}$
plane for chargino masses of
$m_{\tilde{\chi}^{\pm}_1}=50-90$ GeV.
\footnote{We have neglected the diagrams involving virtual sneutrinos,
and assume that
$\sigma (e^+e^-\rightarrow \tilde{\chi}^+_1 \tilde{\chi}^-_1)\approx$
$\sigma (e^+e^-\rightarrow Z^{\ast},\gamma^{\ast} \rightarrow
\tilde{\chi}^+_1 \tilde{\chi}^-_1)$.
This should be a good approximation provided the
sneutrino mass exceeds 500 GeV.
Also we have assumed that the SUSY breaking scale
is equal to $m_Z$ which is consistent for gluino masses of
order 100 GeV.} In each of the figures, we have
fixed the chargino mass, and plotted the cross-section
as a function of the lightest neutralino, for an allowed
range of gluino masses.
These plots clearly show that the LEPII measurements
of the lightest chargino and neutralino masses
(and to a lesser extent the cross-section) enable the gluino
mass to be predicted from the theory. Furthermore the values
of $\tan \beta$ and $\mu$ given at the end-points of the
gluino mass contours will vary along the contour, thus enabling
these parameters to be pinned down by an accurate measurement
of the cross-section. The precise accuracy required is clear
from these figures. The finite length of the contours in Figs.\ 3-7
is due to the Z-pole constraints discussed earlier,
and so for example $\tan \beta$ will vary along each of the contours
up to its maximum permitted value. In some of the contours it will
be noticed that there is a small gap. This gap corresponds to
two different allowed regions with different signs of $\mu$.
These broken contours occur for heavier values of gluino mass
and generally have a hairpin shape. By contrast the contours for
lighter gluino masses are approximately vertical
straight lines, and only have an allowed region for negative $\mu$.
The broken-hairpin contours tend to intersect other contours
leading to ambiguities in the determination of $\mu$, $\tan \beta$ and
$m_{\tilde{g}}$ which may be resolved by a direct measurement of the gluino
mass, for example.

Corresponding to Figs.\ 3-7, Tables 1-5 show the ranges
of $\tan \beta$ and $\mu$ for each of the contours.
In these Tables we have also included the corresponding
ranges of the cross-section and lightest neutralino mass
which will enable the entries in Tables 1-5 to be identified
with the contours in Figs.\ 3-7. The region where the contours are
approximately vertical is particularly interesting. A measurement
of the lightest chargino and neutralino masses in that region
predicts the gluino mass independently of the
value of the total cross section. This prediction is therefore
independent of the sneutrino mass, which we have neglected.
This can be understood
in the following way: if the sneutrino is lighter than about 500 GeV,
its contribution to the total cross section starts to become
non-neglegible and the value of $\sigma(e^+e^-\rightarrow
{\tilde\chi}^+_1{\tilde\chi}^-_1)$ will change with
respect to our approximation,
but the chargino/neutralino spectrum will be unaffected.
This implies that, although
the vertical contours move up and down the vertical axis
when the sneutrino mass is changed, the prediction of the
gluino mass remains unchanged.
Another property of this region is that the
value of the parameter $\tan\beta$ is confined to a narrow
interval close to the unity. This type of upper bound on
$\tan\beta$ is precisely what we have in Fig.\ 2.

We hope that the combination
of Figs.\ 3-7 and Tables 1-5 will prove useful in drawing the
first conclusions about the parameters $\mu$, $\tan \beta$ and
$m_{\tilde{g}}$ from the experimental measurements
of $\sigma (e^+e^-\rightarrow \tilde{\chi}^+_1 \tilde{\chi}^-_1)$,
$m_{\tilde{\chi}^{0}_1}$ and $m_{\tilde{\chi}^{\pm}_1}$.
With knowledge of $\mu$, $\tan \beta$ and
$m_{\tilde{g}}$ one may
predict the entire spectrum of chargino and neutralino
masses and mixing angles using standard formulae, and then search
for the remaining particles in this sector.
The particle physics and cosmological implications
can hardly be overstated, and so it is with some excitement that we
await the first results from LEPII.

\newpage

{\footnotesize
Table 1.
The numerical values of $\tan \beta$, $\mu$ (GeV),
$\sigma (e^+e^-\rightarrow \tilde{\chi}^+_1 \tilde{\chi}^-_1)$ (pb),
$m_{\tilde{\chi}^{0}_1}$ (GeV) at the end-points of the
gluino contours in Fig.\ 3 corresponding to
$m_{\tilde{\chi}^{\pm}_1}=$50 GeV.}

\begin{tabular}{|c|c|c|c|c|} \hline
$m_{\tilde{g}}$ & $\tan \beta$ & $\mu$ &
$\sigma$ &
$m_{\tilde{\chi}^{0}_1}$ \\ \hline
100 & $1.00\rightarrow1.62$ & $-270 \rightarrow-227 $ &
      $7.48\rightarrow7.35$ & $17.9$ \\ \hline
110 & $1.00\rightarrow1.75$ & $-326 \rightarrow-263 $ &
      $7.58\rightarrow7.45$ & $19.4\rightarrow19.5$ \\ \hline
120 & $1.00\rightarrow1.95$ & $-406 \rightarrow-303 $ &
      $7.67\rightarrow7.53$ & $20.9\rightarrow21.0$ \\ \hline
130 & $1.00\rightarrow2.30$ & $-528 \rightarrow-348 $ &
      $7.74\rightarrow7.60$ & $22.3$ \\ \hline
140 & $1.00\rightarrow2.87$ & $-741 \rightarrow-400 $ &
      $7.80\rightarrow7.65$ & $23.5\rightarrow23.6$ \\ \hline
160 & $5.03\rightarrow7.40$ & $-991 \rightarrow-503 $ &
      $7.82\rightarrow7.71$ & $25.6\rightarrow25.7$ \\ \hline
180 & $3.42\rightarrow9.91$ & $995  \rightarrow506  $ &
      $7.82\rightarrow7.70$ & $26.8\rightarrow27.0$ \\ \hline
200 & $1.00\rightarrow3.07$ & $707  \rightarrow462  $ &
      $7.77\rightarrow7.65$ & $27.3\rightarrow27.5$ \\ \hline
220 & $1.00\rightarrow2.04$ & $458  \rightarrow382  $ &
      $7.62\rightarrow7.53$ & $27.2\rightarrow27.4$ \\ \hline
250 & $1.00\rightarrow1.69$ & $310  \rightarrow282  $ &
      $7.31\rightarrow7.21$ & $26.5\rightarrow26.8$ \\ \hline
300 & $1.00\rightarrow22.9$ & $212  \rightarrow100  $ &
      $6.64\rightarrow5.33$ & $25.2\rightarrow31.0$ \\ \hline
400 & $1.00\rightarrow4.57$ & $142  \rightarrow100  $ &
      $5.40\rightarrow4.82$ & $24.7\rightarrow29.6$ \\ \hline
600 & $1.00\rightarrow4.37$ & $99.7 \rightarrow77.0 $ &
      $4.14\rightarrow3.93$ & $28.3\rightarrow32.2$ \\ \hline
800 & $1.00\rightarrow100$ & $84.0\rightarrow56.1$ &
      $3.71\rightarrow3.57$ &
      $32.5\rightarrow35.3\rightarrow35.2$ \\ \hline
 "  & $50.1\rightarrow2.00$ & $-54.4\rightarrow-33.2$ &
      $3.56\rightarrow3.48$ & $35.1\rightarrow30.3$ \\ \hline
1000 & $1.00\rightarrow 100$ & $75.8\rightarrow54.0$ &
       $3.52\rightarrow3.45$ &
       $35.8\rightarrow37.3\rightarrow37.1$ \\ \hline
 "   & $100\rightarrow1.74$ & $-53.1\rightarrow-34.5$ &
       $3.45\rightarrow3.40$ & $37.0\rightarrow32.8$ \\ \hline
\end{tabular}

\vspace{0.25in}
\newpage

{\footnotesize
Table 2.
The numerical values of $\tan \beta$, $\mu$ (GeV),
$\sigma (e^+e^-\rightarrow \tilde{\chi}^+_1 \tilde{\chi}^-_1)$ (pb),
$m_{\tilde{\chi}^{0}_1}$ (GeV) at the end-points of the
gluino contours in Fig.\ 4 corresponding to
$m_{\tilde{\chi}^{\pm}_1}=$60 GeV.}

\begin{tabular}{|c|c|c|c|c|} \hline
$m_{\tilde{g}}$ & $\tan \beta$ & $\mu$ &
$\sigma$ &
$m_{\tilde{\chi}^{0}_1}$ \\ \hline
100 & $1.00\rightarrow1.46$ & $-154 \rightarrow -130$ &
      $6.86\rightarrow6.67$ & $18.2$ \\ \hline
110 & $1.00\rightarrow1.57$ & $-177 \rightarrow -143$ &
      $6.99\rightarrow6.75$ & $19.9$ \\ \hline
120 & $1.00\rightarrow1.69$ & $-207 \rightarrow -158$ &
      $7.10\rightarrow6.83$ & $21.5\rightarrow21.6$ \\ \hline
130 & $1.00\rightarrow1.87$ & $-245 \rightarrow -171$ &
      $7.21\rightarrow6.89$ & $23.1\rightarrow23.2$ \\ \hline
140 & $1.00\rightarrow2.08$ & $-295 \rightarrow -185$ &
      $7.31\rightarrow6.95$ & $24.6\rightarrow24.7$ \\ \hline
160 & $1.00\rightarrow2.85$ & $-469 \rightarrow -193$ &
      $7.48\rightarrow6.94$ & $27.3\rightarrow27.6$ \\ \hline
180 & $1.00\rightarrow4.07$ & $-981 \rightarrow -276$ &
      $7.58\rightarrow7.20$ & $29.6\rightarrow29.9$ \\ \hline
200 & $21.9\rightarrow5.25$ & $-985\rightarrow-145$ &
      $7.58\rightarrow6.40$ & $31.3\rightarrow32.8$ \\ \hline
220 & $1.91\rightarrow 100$ & $987\rightarrow271$ &
      $7.58\rightarrow7.09$ & $32.4\rightarrow33.2$ \\ \hline
  " & $100 \rightarrow5.50$ & $-248\rightarrow-115$ &
      $7.01\rightarrow5.90$ & $33.3\rightarrow35.1$ \\ \hline
250 & $1.00\rightarrow 100$ & $496\rightarrow169$ &
      $7.41\rightarrow6.38$ & $32.9\rightarrow35.5$ \\ \hline
 "  & $100 \rightarrow4.79$ & $-159\rightarrow-84.7$ &
      $6.29\rightarrow5.21$ & $35.7\rightarrow37.6$ \\ \hline
300 & $1.00\rightarrow 100$ & $277\rightarrow121$ &
      $6.88\rightarrow5.51$ & $32.5\rightarrow37.7$ \\ \hline
 "  & $100 \rightarrow4.37$ & $-116\rightarrow-67.8$ &
      $5.43\rightarrow4.61$ &
      $37.9\rightarrow39.0\rightarrow38.4$ \\ \hline
400 & $1.00\rightarrow 100$ & $168\rightarrow90.5$ &
      $5.59\rightarrow4.50$ & $32.2\rightarrow39.3$ \\ \hline
 "  & $100 \rightarrow1.00$ & $-87.6\rightarrow-24.2$ &
      $4.46\rightarrow3.63$ &
      $39.5\rightarrow39.7\rightarrow24.2$ \\ \hline
600 & $1.00\rightarrow 100$ & $114\rightarrow73.1$ &
      $4.13\rightarrow3.73$ & $36.2\rightarrow41.8$ \\ \hline
 "  & $100\rightarrow 1.00$ & $-71.4\rightarrow-33.1$ &
      $3.72\rightarrow3.44$ & $41.8\rightarrow33.1$ \\ \hline
800 & $1.00\rightarrow 100$ & $95.9\rightarrow67.4$ &
      $3.64\rightarrow3.47$ & $41.0\rightarrow44.3$ \\ \hline
 "  & $100 \rightarrow1.00$ & $-66.2\rightarrow-38.5$ &
      $3.46\rightarrow3.34$ & $44.3\rightarrow38.5$ \\ \hline
1000 & $1.00\rightarrow 100$ & $86.9\rightarrow64.8$ &
       $3.44\rightarrow3.35$ & $44.7\rightarrow46.5$ \\ \hline
 "   & $100 \rightarrow1.00$ & $-63.9\rightarrow-42.1$ &
       $3.35\rightarrow3.28$ & $46.4\rightarrow42.1$ \\ \hline
\end{tabular}

\newpage

{\footnotesize
Table 3.
The numerical values of $\tan \beta$, $\mu$ (GeV),
$\sigma (e^+e^-\rightarrow \tilde{\chi}^+_1 \tilde{\chi}^-_1)$ (pb),
$m_{\tilde{\chi}^{0}_1}$ (GeV) at the end-points of the
gluino contours in Fig.\ 5 corresponding to
$m_{\tilde{\chi}^{\pm}_1}=$70 GeV.}

\begin{tabular}{|c|c|c|c|c|} \hline
$m_{\tilde{g}}$ & $\tan \beta$ & $\mu$ &
$\sigma$ &
$m_{\tilde{\chi}^{0}_1}$ \\ \hline
100 & $1.00\rightarrow1.27$ & $-90.5\rightarrow-80.3$ &
      $6.05\rightarrow5.89$ & $18.4$ \\ \hline
110 & $1.00\rightarrow1.38$ & $-103 \rightarrow-84.0$ &
      $6.17\rightarrow5.89$ & $20.2$ \\ \hline
120 & $1.00\rightarrow1.49$ & $-119 \rightarrow-86.0$ &
      $6.30\rightarrow5.87$ & $21.9$ \\ \hline
130 & $1.00\rightarrow1.60$ & $-137 \rightarrow-89.4$ &
      $6.42\rightarrow5.86$ & $23.5\rightarrow23.7$ \\ \hline
140 & $1.00\rightarrow1.73$ & $-159 \rightarrow-89.1$ &
      $6.54\rightarrow5.79$ & $25.1\rightarrow25.4$ \\ \hline
160 & $1.00\rightarrow1.98$ & $-220 \rightarrow -103$ &
      $6.76\rightarrow5.89$ & $28.2\rightarrow28.6$ \\ \hline
180 & $1.00\rightarrow2.31\rightarrow2.20$ &
      $-328 \rightarrow-84.3$ &
      $6.95\rightarrow5.42$ & $31.0\rightarrow31.9$ \\ \hline
200 & $1.00\rightarrow2.95\rightarrow2.40$ &
      $-561 \rightarrow-81.4$ &
      $7.08\rightarrow5.22$ & $33.4\rightarrow35.0$ \\ \hline
220 & $2.34\rightarrow4.57\rightarrow2.40$ &
      $-999\rightarrow-70.2$ &
      $7.14\rightarrow4.83$ & $35.4\rightarrow37.9$ \\ \hline
250 & $2.63\rightarrow100$ & $1000\rightarrow325$ &
      $7.14\rightarrow6.81$ & $37.5\rightarrow38.1$ \\ \hline
 "  & $100\rightarrow2.40$ & $-297\rightarrow-62.7$ &
      $6.75\rightarrow4.45$ &
      $38.3\rightarrow41.3\rightarrow41.2$ \\ \hline
300 & $1.00\rightarrow100$ & $397\rightarrow162$ &
      $6.84\rightarrow5.77$ & $38.7\rightarrow42.1$ \\ \hline
 "  & $100\rightarrow1.66$ & $-154\rightarrow-40.2$ &
      $5.69\rightarrow3.72$ &
      $42.4\rightarrow44.9\rightarrow36.5$ \\ \hline
400 & $1.00\rightarrow100$ & $200\rightarrow109$ &
      $5.65\rightarrow4.49$ & $39.1\rightarrow46.2$ \\ \hline
 "  & $100\rightarrow1.00$ & $-106\rightarrow-36.1$ &
      $4.44\rightarrow3.38$ &
      $46.4\rightarrow47.6\rightarrow36.1$ \\ \hline
600 & $1.00\rightarrow100$ & $129\rightarrow85.8$ &
      $4.03\rightarrow3.57$ & $43.7\rightarrow50.2$ \\ \hline
 "  & $100\rightarrow1.00$ & $-84.1\rightarrow-44.2$ &
      $3.55\rightarrow3.22$ &
      $50.3\rightarrow50.6\rightarrow44.2$ \\ \hline
800 & $1.00\rightarrow100$ & $108\rightarrow78.7$ &
      $3.47\rightarrow3.28$ & $49.3\rightarrow53.4$ \\ \hline
 "  & $100\rightarrow1.00$ & $-77.5\rightarrow-49.2$ &
      $3.28\rightarrow3.13$ & $53.4\rightarrow49.2$ \\ \hline
1000 & $1.00\rightarrow100$ & $98.1\rightarrow75.6$ &
       $3.26\rightarrow3.17$ & $53.5\rightarrow55.9$ \\ \hline
 "   & $100\rightarrow1.00$ & $-74.7\rightarrow-52.6$ &
       $3.16\rightarrow3.09$ & $55.9\rightarrow52.6$ \\ \hline
\end{tabular}

\newpage

\vspace{0.25in}

{\footnotesize
Table 4.
The numerical values of $\tan \beta$, $\mu$ (GeV),
$\sigma (e^+e^-\rightarrow \tilde{\chi}^+_1 \tilde{\chi}^-_1)$ (pb),
$m_{\tilde{\chi}^{0}_1}$ (GeV) at the end-points of the
gluino contours in Fig.\ 6 corresponding to
$m_{\tilde{\chi}^{\pm}_1}=$80 GeV.}

\begin{tabular}{|c|c|c|c|c|} \hline
$m_{\tilde{g}}$ & $\tan \beta$ & $\mu$ &
$\sigma$ &
$m_{\tilde{\chi}^{0}_1}$ \\ \hline
123 & $1.00\rightarrow1.06$ & $-68.9\rightarrow-68.1$ &
      $5.27\rightarrow5.25$ & $22.6$ \\ \hline
130 & $1.00\rightarrow1.19$ & $-76.5\rightarrow-68.1$ &
      $5.35\rightarrow5.17$ & $23.8\rightarrow23.9$ \\ \hline
140 & $1.00\rightarrow1.29$ & $-88.7\rightarrow-69.3$ &
      $5.46\rightarrow5.09$ & $25.5\rightarrow25.6$ \\ \hline
160 & $1.00\rightarrow1.45$ & $-120\rightarrow-76.1$ &
      $5.68\rightarrow5.03$ & $28.8\rightarrow29.0$ \\ \hline
180 & $1.00\rightarrow1.58$ & $-166\rightarrow-66.4$ &
      $5.89\rightarrow4.58$ & $31.9\rightarrow32.5$ \\ \hline
200 & $1.00\rightarrow1.88\rightarrow1.66$ &
      $-239\rightarrow-63.8$ &
      $6.08\rightarrow4.30$ & $34.7\rightarrow35.9$ \\ \hline
220 & $1.00\rightarrow2.19\rightarrow1.74$ &
      $-373\rightarrow-63.6$ &
      $6.23\rightarrow4.12$ & $37.3\rightarrow39.0$ \\ \hline
250 & $1.65\rightarrow3.84\rightarrow1.34$ &
      $-996\rightarrow-44.1$ & $6.36\rightarrow3.34$ &
      $40.4\rightarrow43.5\rightarrow41.8$ \\ \hline
300 & $1.00\rightarrow100$ & $743\rightarrow245$ &
      $6.32\rightarrow5.78$ & $43.5\rightarrow45.0$ \\ \hline
 "  & $100\rightarrow1.00$ & $-231\rightarrow-42.1$ &
      $5.72\rightarrow3.09$ &
      $45.1\rightarrow49.1\rightarrow42.1$ \\ \hline
400 & $1.00\rightarrow100$ & $243\rightarrow132$ &
      $5.41\rightarrow4.32$ & $45.4\rightarrow51.7$ \\ \hline
 "  & $100\rightarrow1.00$ & $-128\rightarrow-47.8$ &
      $4.26\rightarrow2.98$ &
      $52.0\rightarrow54.8\rightarrow47.8$ \\ \hline
600 & $1.00\rightarrow100$ & $145\rightarrow98.8$ &
      $3.74\rightarrow3.24$ & $50.8\rightarrow58.2$ \\ \hline
 "  & $100\rightarrow1.00$ & $-97.1\rightarrow-55.2$ &
      $3.23\rightarrow2.85$ &
      $58.4\rightarrow59.3\rightarrow55.2$ \\ \hline
800 & $1.00\rightarrow100$ & $120\rightarrow90.2$ &
      $3.15\rightarrow2.95$ & $57.2\rightarrow62.3$ \\ \hline
 "  & $100\rightarrow1.00$ & $-89.0\rightarrow-59.8$ &
      $2.94\rightarrow2.78$ &
      $62.3\rightarrow62.6\rightarrow59.8$ \\ \hline
1000 & $1.00\rightarrow100$ & $109\rightarrow86.4$ &
       $2.92\rightarrow2.83$ & $62.2\rightarrow65.2$ \\ \hline
 "   & $100\rightarrow1.00$ & $-85.5\rightarrow-63.0$ &
       $2.82\rightarrow2.75$ & $65.2\rightarrow63.0$ \\ \hline
\end{tabular}


{\footnotesize
Table 5.
The numerical values of $\tan \beta$, $\mu$ (GeV),
$\sigma (e^+e^-\rightarrow \tilde{\chi}^+_1 \tilde{\chi}^-_1)$ (pb),
$m_{\tilde{\chi}^{0}_1}$ (GeV) at the end-points of the
gluino contours in Fig.\ 6 corresponding to
$m_{\tilde{\chi}^{\pm}_1}=$90 GeV.}

\begin{tabular}{|c|c|c|c|c|} \hline
$m_{\tilde{g}}$ & $\tan \beta$ & $\mu$ &
$\sigma$ &
$m_{\tilde{\chi}^{0}_1}$ \\ \hline
160 & $1.00\rightarrow1.05$ & $-62.6\rightarrow-61.0$ &
      $4.11\rightarrow4.05$ & $29.3$ \\ \hline
180 & $1.00\rightarrow1.21$ & $-87.9\rightarrow-70.8$ &
      $4.28\rightarrow3.93$ & $32.5\rightarrow32.7$ \\ \hline
200 & $1.00\rightarrow1.36\rightarrow1.24$ &
      $-123\rightarrow-56.5$ &
      $4.45\rightarrow3.05$ & $35.6\rightarrow36.3$ \\ \hline
220 & $1.00\rightarrow1.52\rightarrow1.23$ &
      $-176\rightarrow-54.8$ &
      $4.61\rightarrow2.75$ & $38.5\rightarrow39.7$ \\ \hline
250 & $1.00\rightarrow2.00\rightarrow1.00$ &
      $-332\rightarrow-50.9$ &
      $4.80\rightarrow2.40$ & $42.3\rightarrow45.0$ \\ \hline
300 & $14.5\rightarrow1.00$ & $-962\rightarrow-54.2$ &
      $4.92\rightarrow2.36$ & $47.0\rightarrow53.4$ \\ \hline
400 & $1.00\rightarrow100$ & $307\rightarrow160$ &
      $4.47\rightarrow3.67$ & $50.9\rightarrow55.7$ \\ \hline
 "  & $100\rightarrow1.00$ & $-155\rightarrow-59.3$ &
      $3.62\rightarrow2.28$ &
      $56.0\rightarrow61.3\rightarrow59.3$ \\ \hline
600 & $1.00\rightarrow100$ & $162\rightarrow112$ &
      $3.03\rightarrow2.57$ & $57.4\rightarrow65.5$ \\ \hline
 "  & $100\rightarrow1.00$ & $-111\rightarrow-66.1$ &
      $2.56\rightarrow2.19$ &
      $65.8\rightarrow68.1\rightarrow66.1$ \\ \hline
800 & $1.00\rightarrow100$ & $133\rightarrow102$ &
      $2.48\rightarrow2.30$ & $64.9\rightarrow70.9$ \\ \hline
 "  & $100\rightarrow1.00$ & $-100\rightarrow-70.5$ &
      $2.29\rightarrow2.15$ &
      $71.0\rightarrow71.9\rightarrow70.5$ \\ \hline
1000 & $1.00\rightarrow100$ & $121\rightarrow97.3$ &
       $2.28\rightarrow2.19$ & $70.6\rightarrow74.4$ \\ \hline
 "   & $100\rightarrow1.00$ & $-96.4\rightarrow-73.5$ &
       $2.19\rightarrow2.12$ &
       $74.4\rightarrow74.7\rightarrow73.5$ \\ \hline
\end{tabular}

\newpage

\noindent
{\bf \large Figure Captions:}

\noindent
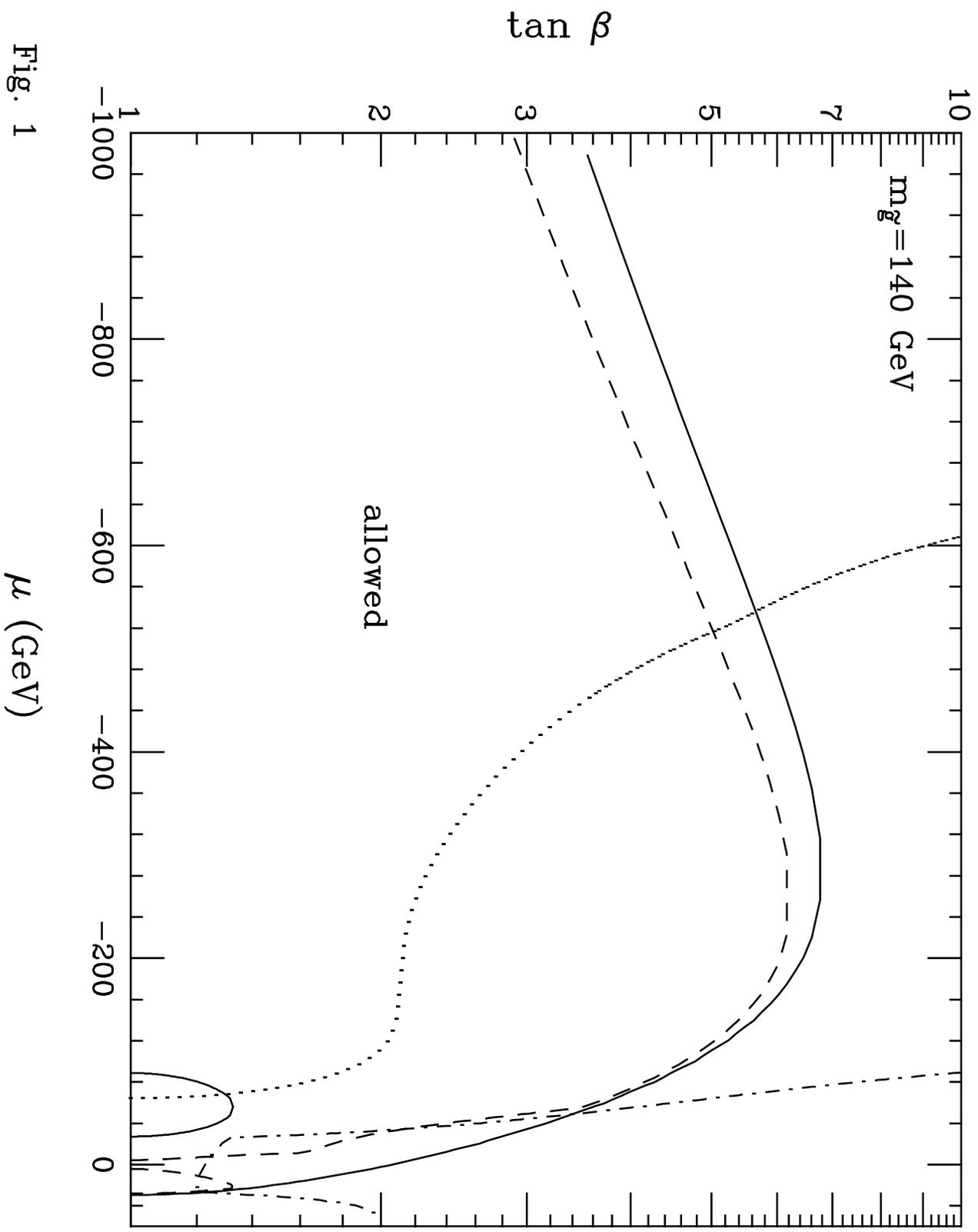
{\bf Fig. 1.}
Allowed region in the $\tan\beta-\mu$ plane for $m_{\tilde g}=140$
GeV. The different regions below each curve correspond to the
following constraints: $m_{\tilde\chi^{\pm}_1}\geq 45$ GeV (large
solid curve), $\Delta\Gamma_Z\leq 30$ MeV (dashes),
$\Delta\Gamma_Z^{inv}\leq 7$ MeV (dotdash), $B(Z\rightarrow
\tilde\chi^0_i\tilde\chi^0_j)\leq 1\times 10^{-5}$ (dots), and,
in the case a new lower bound on the chargino mass is found,
$m_{\tilde\chi^{\pm}_1}\ge 80$ GeV (small solid curve).
\\
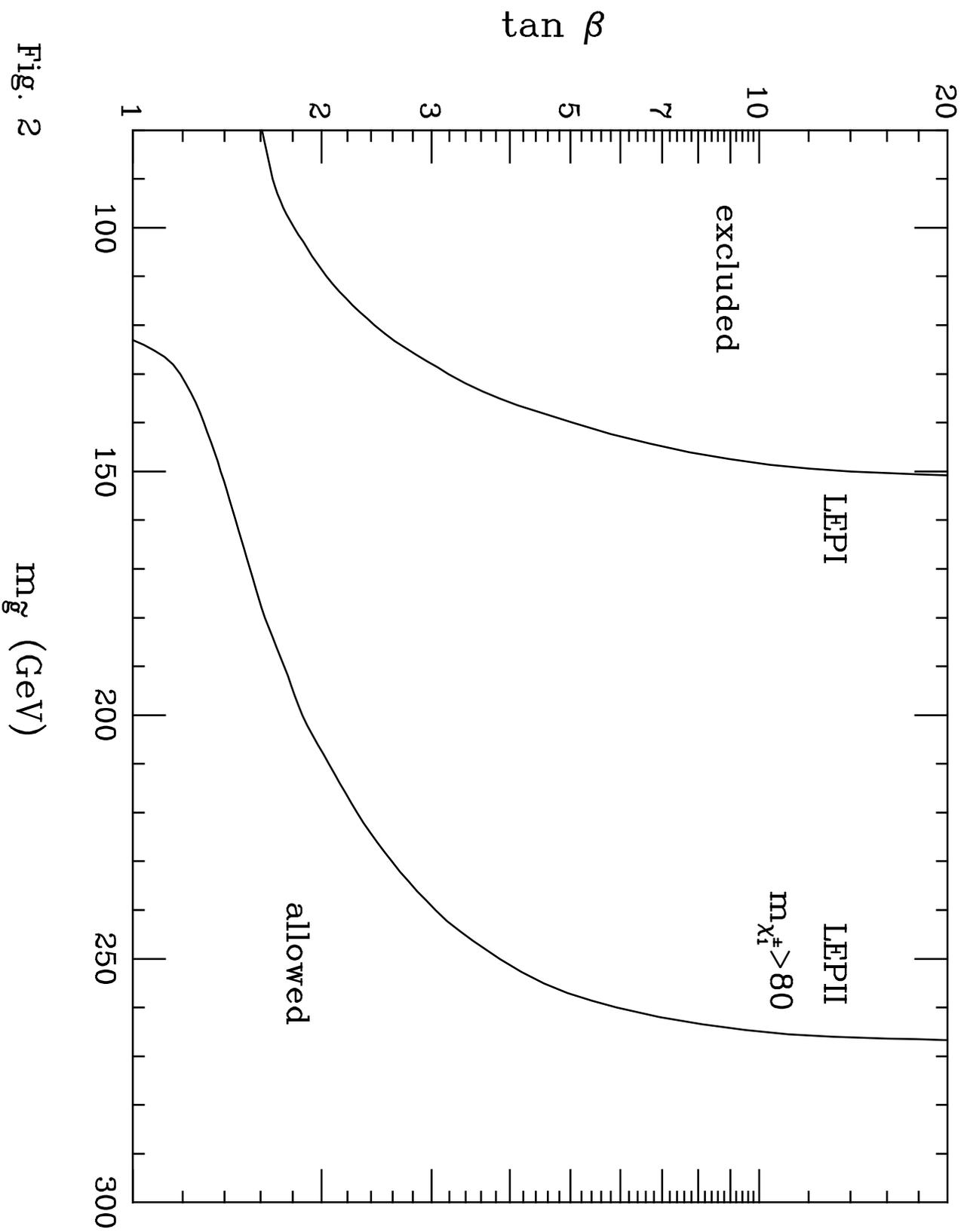
{\bf Fig. 2.}
Gluino mass dependent upper bound on $\tan\beta$. The curve on the
left corresponds to the LEP data, and the curve on the right
corresponds to a hypothetical new lower bound on the chargino
mass ($m_{\tilde\chi^{\pm}_1}>80$ GeV) at LEPII. In each case,
the allowed region lies below and at the right of the curve.
The experimental lower bound on the gluino mass is 100 GeV,
with the exception of the light gluino window (see the text).
\\
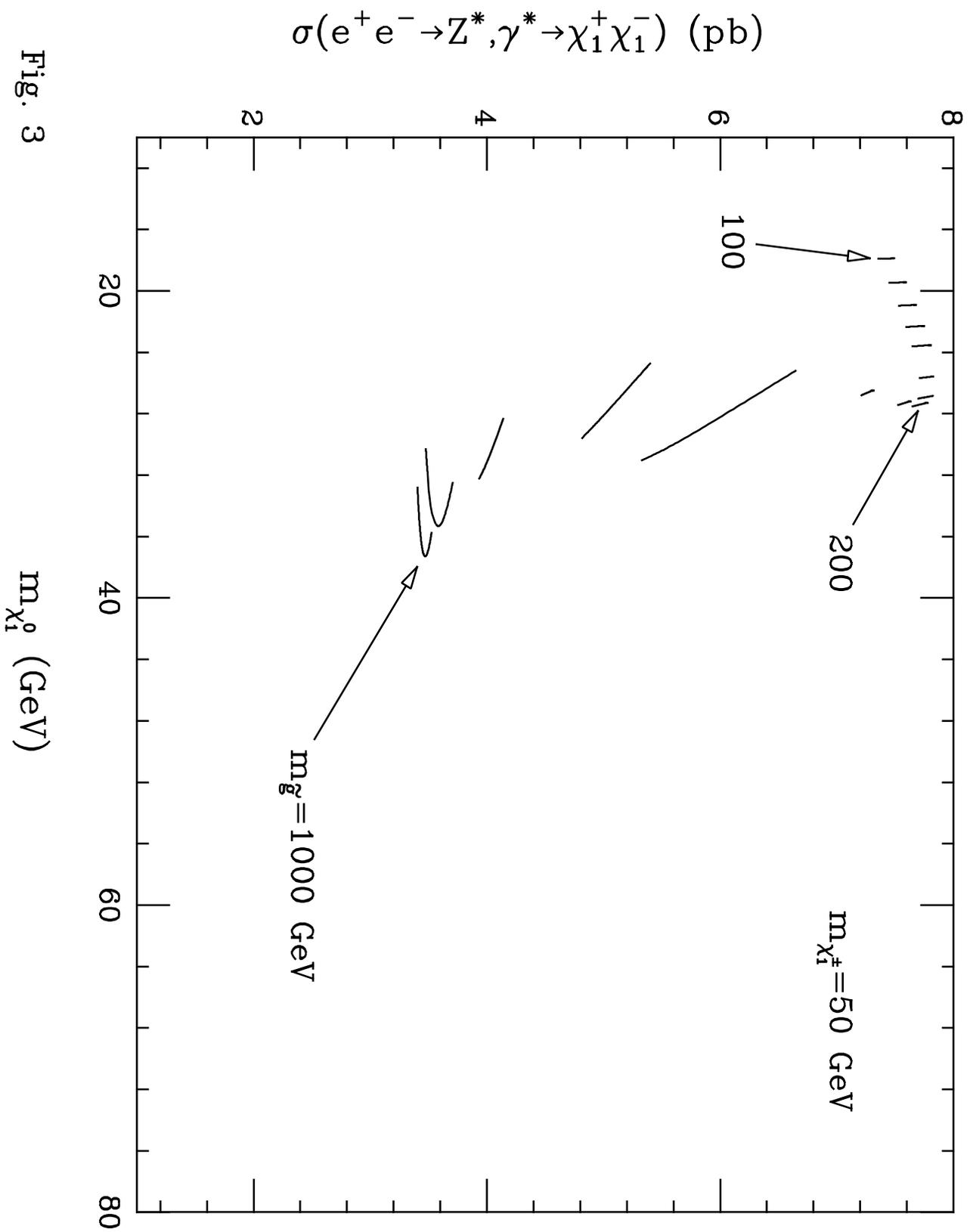
{\bf Fig. 3.}
For a chargino mass given by $m_{\tilde\chi^{\pm}_1}=50$ GeV,
we plot contours of equal gluino mass $m_{\tilde g}=$100, 110,
120, 130, 140, 160, 180, 200, 220, 250, 300, 400, 600, 800,
1000 GeV, in the plane $\sigma(e^+e^-\rightarrow Z^*,\gamma^*
\rightarrow\chi^+_1\chi^-_1)-m_{\chi^0_1}$.
\\
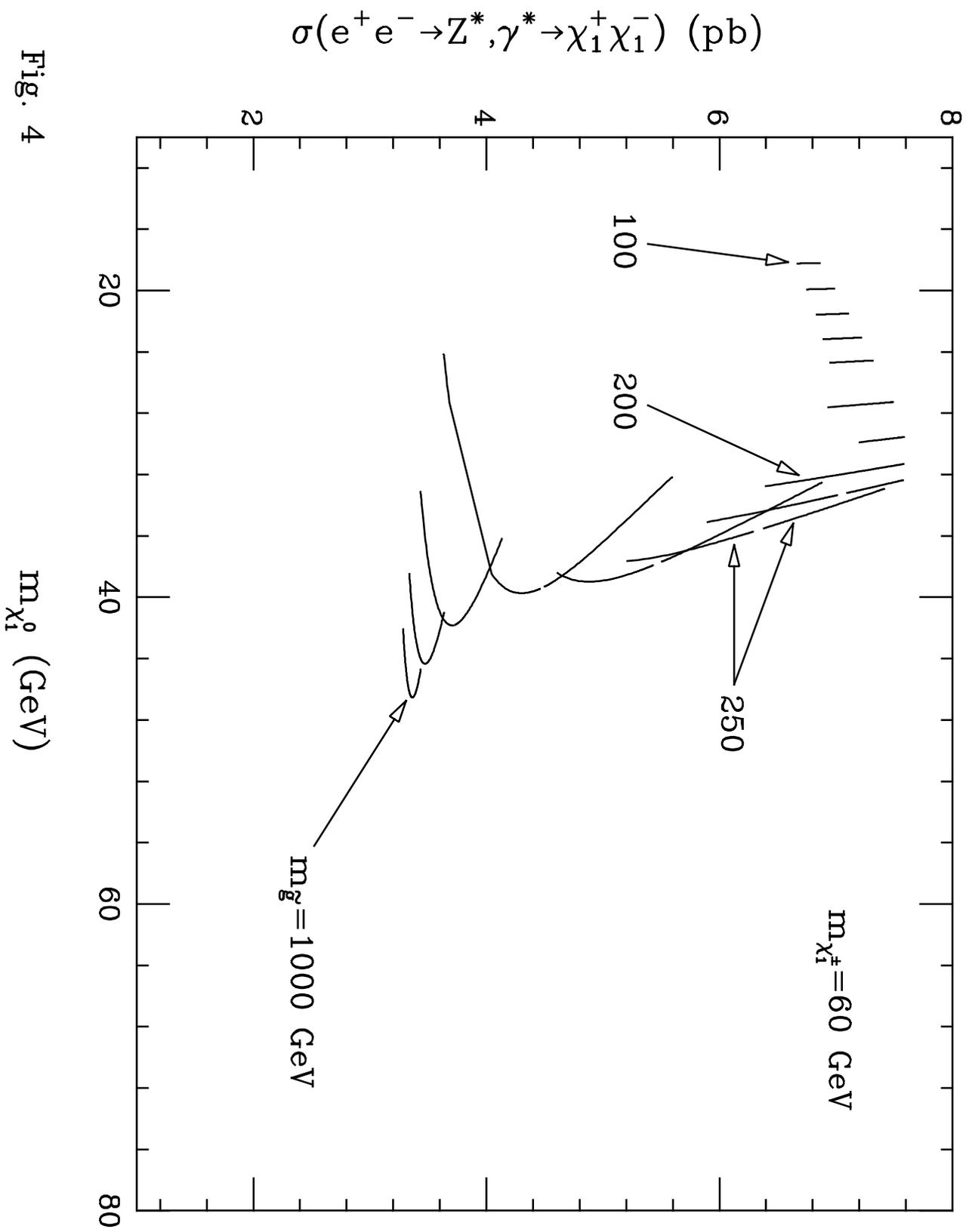
{\bf Fig. 4.}
Same as Fig. 3 with $m_{\tilde\chi^{\pm}_1}=60$ GeV.
\\
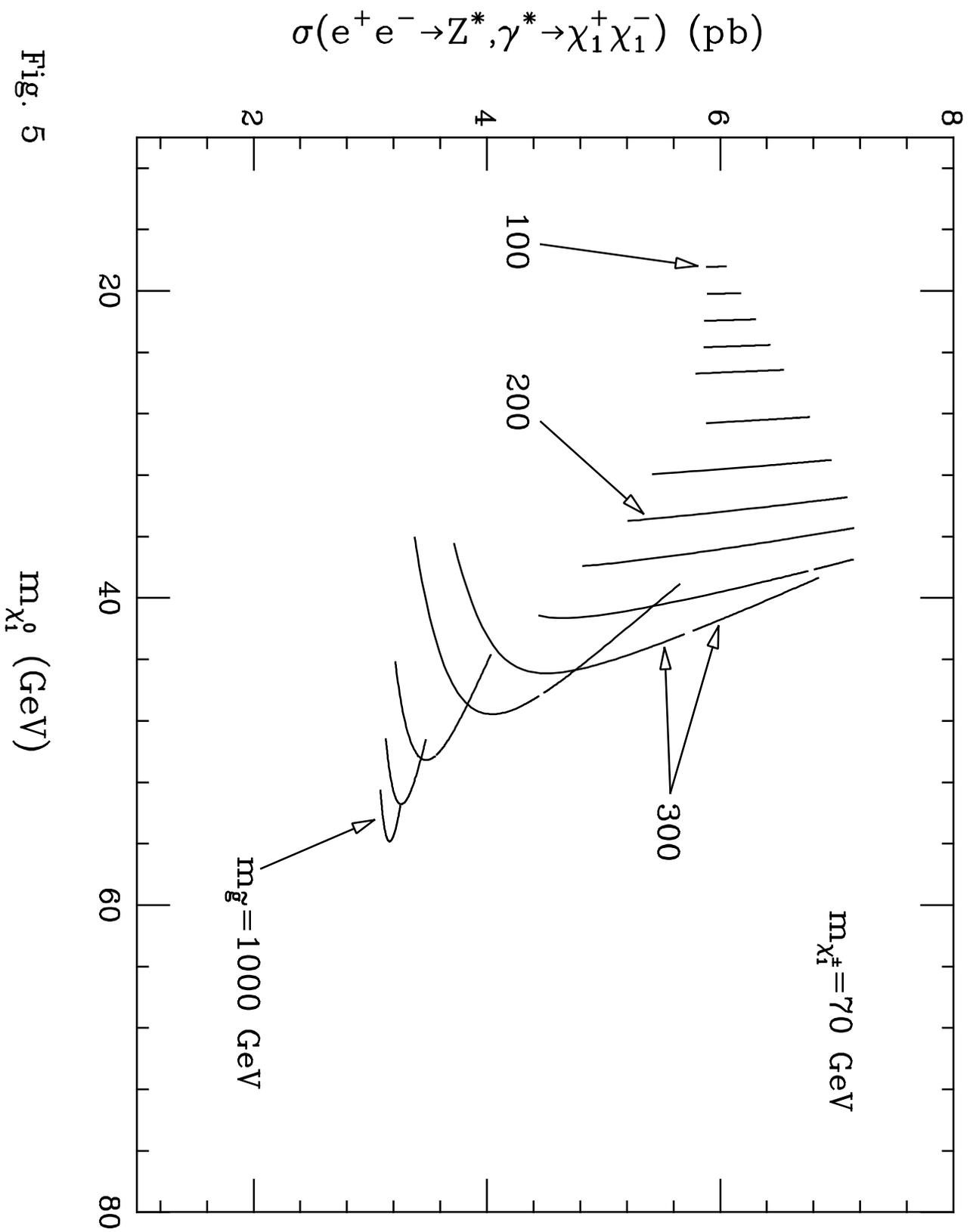
{\bf Fig. 5.}
Same as Fig. 3 with $m_{\tilde\chi^{\pm}_1}=70$ GeV.
\\
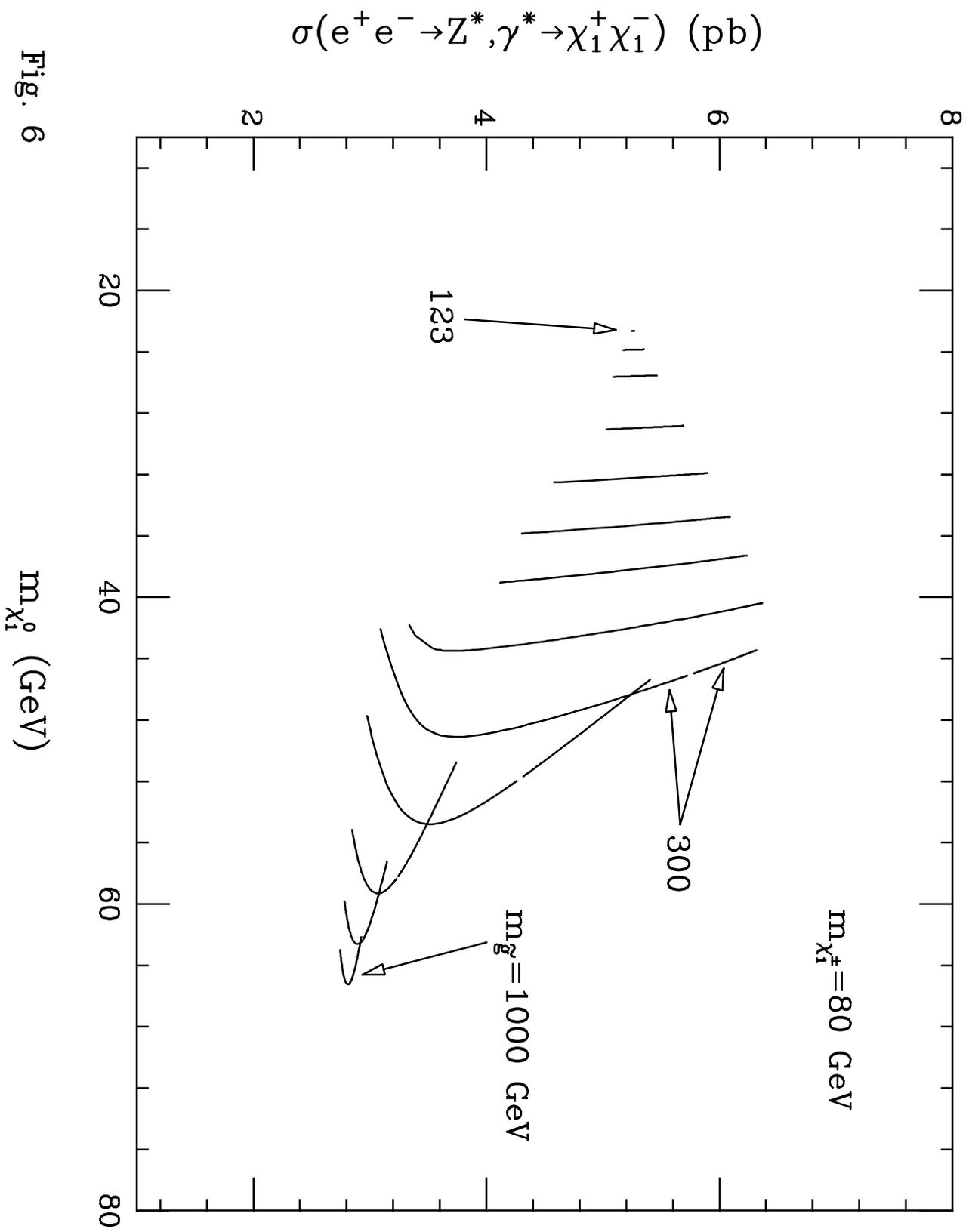
{\bf Fig. 6.}
Same as Fig. 3 with $m_{\tilde\chi^{\pm}_1}=80$ GeV
and for gluino masses given by $m_{\tilde g}=$123, 130,
140, 160, 180, 200, 220, 250, 300, 400, 600, 800, 1000
GeV.
\\
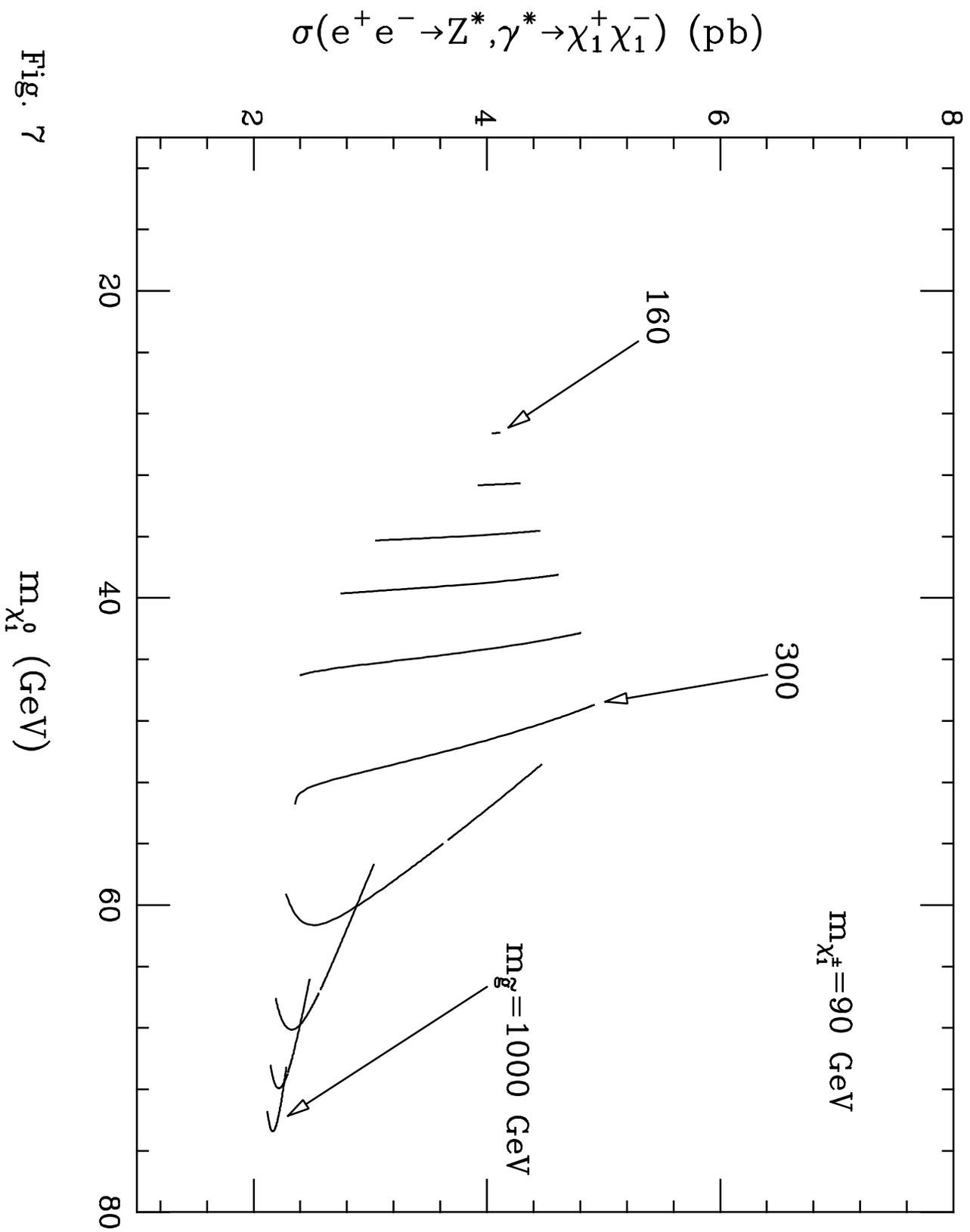
{\bf Fig. 7.}
Same as Fig. 3 with $m_{\tilde\chi^{\pm}_1}=90$ GeV
and for gluino masses given by $m_{\tilde g}=$160, 180,
200, 220, 250, 300, 400, 600, 800, 1000 GeV.
\\

\end{document}